\documentstyle[aps,prb]{revtex}
\begin{document}
\newcommand{\be}{\begin{equation}}
\newcommand{\ee}{\end{equation}}
\newcommand{\ba}{\begin{eqnarray}}
\newcommand{\ea}{\end{eqnarray}}
\title{Anomalous Tunneling Conductances of a
Spin Singlet $\nu=2/3$ Edge States: Interplay of Zeeman Splitting and Long Range 
Coulomb Interaction} 
\author{H. C. Lee} 
\address{ Asia
Pacific Center for Theoretical Physics, Seoul, Korea  } 
\draft
\maketitle 
\begin{abstract}
\indent The point contact tunneling conductance
between edges of the spin singlet $\nu=2/3,\hat{K}=(3/3/0)$ quantum Hall states 
is studied both in the quasiparticle tunneling picture 
and in the electron tunneling picture..
Due to the interplay of Zeeman splitting and the long range Coulomb interaction between edges of 
opposite chirality  novel spin
excitations emerge, and their effect is characterized by anomalous exponents of the charge and spin 
tunneling conductances in various temperature ranges. 
Depending on the kinds of scatterings at the point contact and the tunneling mechanism 
the anomalous interaction in spin sector  may enhance or suppress the tunneling conductances. 
The effects of novel spin excitation are
 also relevant to the recent NMR experiments on quantum Hall edges.
\end{abstract}
\pacs{{\rm PACS numbers}: \hspace{.05in} 73.20.Dx, 73.20.Mf, 73.40.Hm } 
\section{Introduction}
Edges of quantum Hall (QH) bar provide many clean experimental verifications of the anomalous properties
of one dimensional system \cite{wen}.  Especially, the power law 
dependences of conductances on temperature and voltage  demonstrate very clearly the (chiral)
Tomonaga-Luttinger liquid character of edge states \cite{experiments}. 
The long range Coulomb interaction (LRCI) between edges of 
opposite chirality  brings new effects into these systems 
\cite{moon,imura2,brey}. The effect of LRCI is manifest in the temperature dependence of the
point contact tunneling conductance of $\nu=\frac{1}{2k +1}$  spin polarized Laughlin states
\cite{moon,imura2}. 
\be
G (T) \sim \cases{ (\frac{T_w}{T} )^{2(1-1/\nu)},\quad \quad T > T_w \cr
       (\frac{T_w}{T} )^{2} \exp \big(-\frac{2 \sqrt{2 \chi}}{3 \nu}
     ( \ln \frac{T_w}{T})^{3/2} \big ),\quad \quad T < T_w, \cr}
\ee
where $T_w$ is a cross-over temperature. The authors of \cite{moon,imura2}  proposed that
the above LRCI effect might explain the discrepancy between experiments and the predictions from  
the exact calculations of the chiral Luttinger liquid (CLL) theory \cite{fendley}.

The spin charge separation is the one of the most prominent features of one dimensional electronic
system. The spin charge separation can be revealed very clearly in the tunneling density of states
of $\nu=2 $ edges \cite{spin-charge}.
In ordinary one dimensional system the LRCI affects only the charge sector  due to the spin
charge separation, and the spin sector remains unchanged \cite{schulz}. 
However, in QH edges or quantum wires in a strong magnetic field the LRCI can influence 
spin sector indirectly \cite{leeyang2}.
In spin singlet edge states, the spin-up and spin-down electrons at Fermi wavevector
are spatially separated due to the Zeeman splitting and the dependence of the guiding center of single
particle wave functions on the wave number. The above separation of spin-up and spin-down edges 
at the Fermi energy induces a nontrivial spin dependence of LRCI between edges of the opposite 
chirality through the cutoff length (or time) scales, and as a result, 
 anomalous spin excitations emerge in the spin sector.  The implication of the spin dependence of LRCI
 has been studied for the spin correlation
functions of $\nu=2$ {\em Fermi Liquid} edge states \cite{leeyang2}.

In this paper, the effect of the anomalous spin excitation in $\nu=2/3$ spin singlet egde are studied
being focused on the temperature dependence of the point contact tunneling conductance. Without LRCI 
$\nu=2/3$ spin singlet state can be described by CLL theory \cite{imura3}.
The ground state of 
$\nu=2/3$ is discovered experiementally \cite{eisenstein} to be spin singlet at low magnetic field and to be
spin polarized at high magnetic field.  There exist two 
spin singlet states\cite{imura3,macdonald} at $\nu=2/3$;  
$\hat{K}=\pmatrix{1 & 2 \cr  2 & 1 \cr}$ and 
$\hat{K}=\pmatrix{3 & 0 \cr 0 & 3 \cr}$ in terms of $K$-matrix \cite{wenzee}.
Both of them are realized in the double layered system, 
and the hopping amplitude and the distance between layers
determine the actual spin singlet ground state \cite{macdonald}.
The $\hat{K}=(1/1/2)$ state can be constructed via the standard hierarchy construction \cite{wen}, and the 
$\hat{K}=(3/3/0)$ state is just an independent $\nu=1/3$ Laughlin state for each spin. In the hierarchy 
spin singlet state, the charge and spin  edge modes propagate in opposite direction, and that makes
the exponent of the tunneling conductance non-universal. To obtain the universal exponent, the
random impurity scatterings at edges are required \cite{kfp}. Since the treatment of LRCI and random impurity 
scattering at the same time is quite complicated,
we  consider the $\hat{K}=(3/3/0)$ state, 
where the random impurity scatterings  at edges are   irrelevant, so that
we can focus on the effect of LRCI between the edges of opposite chirality.

The  effect of the spin dependence of LRCI can be characterized by 
the anomalous exponent $\alpha_\sigma$. 
The anomalous exponent $\alpha_\sigma$ is always
greater than unity, and  in the absence of the spin dependence of LRCI, $\alpha_\sigma=1$. Therefore, the
deviation of $\alpha_\sigma$ from unity is the precise measure of the magnitude of the anomalous interaction in the
spin sector.
The anomalous exponent $\alpha_\sigma$ may enhance or suppress the charge and spin tunneling conductances
depending on if the spin flip is allowed 
at the point contact or not,
 and if the quasiparticle tunneling picture or the electron tunneling picture is used.
For instance, the charge conductance at low temperature within QPT when the spin-flip scattering is present is
given by
\be
G_c(T) \sim  T^7 \times T^{3 (\alpha_\sigma+\alpha_\sigma^{-1}-2)},
\ee
where the subleading term is not shown.
Clearly the anomalous spin excitation {\em  reduces} the charge conductance. 
The spin conductance with the same condition as the charge conductance is given by
\be
G_s(T) \sim T^{10} \times T^{-12 (1-\alpha_\sigma^{-1})}.
\ee
In constrast to the charge conductance, the anomalous spin excitation {\em enhances} the spin conductance.

This paper is organized as follows. In section II, the basic model is introduced,
 and the effective action for the tunneling is derived. The tunneling conductance is calculated in QPT picture
 in section III, and in ET picture in section IV.  We conclude the paper with summary in section V. 

\section{Model} 
The effective action describing two-channel
$\hat{K}=(3/3/0)$ spin singlet edge state with intra- and inter-edge LRCI 
is ($I,J=\uparrow, \downarrow$,$\phi^{\pm}_I=\phi_I^u\pm \phi_I^d \;$) 
\cite{moon,imura2,leeyang2,imura3},
\ba 
S_{{\rm eff}}&=& \int dx d\tau \,\sum_{IJ}\,\Bigg[
\frac{i}{4\pi}\,K_{IJ}\,\partial_\tau \phi^+_I\,\partial_x \phi_J^- +
\frac{1}{8\pi}\, V_{IJ}\, \Big( \partial_x \phi^+_I\,\partial_x \phi^+_J+
\partial_x \phi^-_I\,\partial_x \phi^-_J\Big) \Bigg] \nonumber \\
 &+& \int \,
dx \,dy \,d\tau\sum_{I J}\frac{1}{16 \pi^2}\, \Bigg[
\Big(V_{IJ}^{a}(x-y)+V_{IJ}^{w}(x-y) \Big)\,\partial_x \, \phi^+_I (x) \,
\partial_y \, \phi^+_J (y) \nonumber \\
 &+&\Big(V^a_{IJ}(x-y)-V^w_{IJ}(x-y)
\Big)\,\partial_x \, \phi^-_I (x) \, \partial_y  \, \phi^-_J(y)  \Bigg], \ea
where $V_{IJ}=\pmatrix{v_\uparrow & v_{{\rm int}} \cr v_{{\rm int}} & 
v_\downarrow \cr}$ is the short range interaction between edges with
the same chirality (upper and lower edges), and the intra-edge ($V^a$'s) and inter-edge($V^w$'s)  LRCI are
given by
\ba
V^{a}_{\uparrow \uparrow}(x)&=&V^{a}_{\downarrow \downarrow}(x)=
\frac{e^2}{\epsilon \, \sqrt{x^2+a_{\parallel}^2}},\quad 
V^{a}_{\uparrow \downarrow}(x)=\frac{e^2}{\epsilon \, \sqrt{x^2+a_{\perp}^2}},
\nonumber \\
V^{w}_{\uparrow \uparrow}(x)&=&
\frac{e^2}{\epsilon \, \sqrt{x^2+w_{\parallel \uparrow}^2}},\quad
V^{w}_{\downarrow \downarrow}(x)=
\frac{e^2}{\epsilon \, \sqrt{x^2+w_{\parallel \downarrow}^2}},\quad 
V^{w}_{\uparrow \downarrow}(x)=\frac{e^2}{\epsilon \, \sqrt{x^2+w_{\perp}^2}}.
\ea
$a_\parallel, a_\perp$ are the length scales of order of lattice spacing.
$w_{\parallel s}, w_\perp$ are of the order of the width of Hallbar, and they depend on spin because of
the spatial separation of spin-up and spin-down electrons at the Fermi wavevector \cite{leeyang2,zeeman}.
The above LRCI matrix
elements in momentum space read ($s=\uparrow, \downarrow$),
\be
V^{a}_{k,ss}=
-\frac{ 2 e^2}{\epsilon}\,\ln \frac{\gamma k a_{\parallel}}{2},\;\;
V^{a}_{k,\uparrow \downarrow}=
-\frac{ 2 e^2}{\epsilon}\,\ln \frac{\gamma k a_{\perp}}{2},\;\; 
V^{w}_{k,ss}=
\frac{ 2 e^2}{\epsilon}\,K_0(k w_{\parallel,s}),\;\;
V^{w}_{k,\uparrow \downarrow}=
\frac{ 2 e^2}{\epsilon}\,K_0(k w_{\perp}).
\ee
$K_0(x)$ is the modified Bessel function, and $\gamma=0.5772\cdots$ is the Euler-Mascheroni constant.
For later uses, we define
\ba
v&\equiv&\frac{1}{2}\,\Big(v_{\uparrow}+v_{ \downarrow} \Big),\;\;
V^{a}_{\rho,k}\equiv\frac{1}{2}\Big(V^{a}_{k,ss}+
V^{a}_{k,\uparrow \downarrow}\Big),\;\;
V^{a}_{\sigma,k}\equiv\frac{1}{2}\Big(V^{a}_{k,ss}-
V^{a}_{k,\uparrow \downarrow} \Big),\nonumber \\
V^{w}_{k,\parallel}&\equiv&\frac{1}{2}\,\Big(V^{w}_{k\uparrow \uparrow}
+V^{w}_{k,\downarrow\downarrow}\Big),\;\;
V^{w}_{\rho,k}\equiv\frac{1}{2}\Big(V^{w}_{k,\parallel}+
V^{w}_{k,\uparrow \downarrow}\Big),\;\;
V^{w}_{\sigma,k}\equiv\frac{1}{2}\Big(V^{w}_{k,\parallel}-
V^{w}_{k,\uparrow \downarrow} \Big).
\ea
It is convenient to introduce the following charge-spin basis.
\be
\phi^+_{\uparrow}+\phi^+_{\downarrow}=\phi_c^+,\;\;
\phi^+_{\uparrow}-\phi^+_{\downarrow}=\phi_s^+,\;\;
\phi^-_{\uparrow}+\phi^-_{\downarrow}=\phi_c^-,\;\;
\phi^-_{\uparrow}-\phi^-_{\downarrow}=\phi_s^-.
\ee
The action in momentum-frequency space and in a matrix form \cite{note1} is
\ba
S&=&\frac{1}{16\pi}\,\int \,\frac{d \omega \,d k}{(2\pi)^2}\,\Phi\,
\pmatrix{v_{\rho +} k^2 &  i \omega k K & g_+ k^2  &  0 \cr
          i \omega k K & v_{\rho -} k^2  & 0 & g_- k^2  \cr
	  g_+ k^2 & 0 &  v_\sigma^+ k^2   &  i \omega k K  \cr
	  0  & g_- k^2 &  i \omega k K  & v_\sigma^-  k^2  \cr}\,\Phi^T, \nonumber \\
v_{\rho \pm}&=&v+v_{{\rm int}}+\frac{V^a_{\rho,k}\pm V^w_{\rho,k}}{\pi},\;\;
v_{\sigma \pm}=v-v_{{\rm int}}+\frac{V^a_{\sigma,k}\pm V^w_{\sigma,k}}{\pi} \nonumber \\
g_\pm&=&(v_\uparrow - v_\downarrow)/2\pm 
\frac{1}{2}\,\Big( V^w_{k,\parallel,\uparrow}-V^w_{k,\parallel,\downarrow} \Big),\quad
\Phi=\Big(\phi^+_c,\;\;\phi^-_c,\;\;\phi^+_s,\;\;\phi^-_s \Big).	 
\ea	
Since we are interested in the point contact conductance, the continuum degrees of freedom
can be integrated out, and they act as a reservoir for the degree of freedom at the
point contact.
The effective action of $\theta(\tau) \equiv  \Phi(x=0, \tau)$ is
\cite{kanefisher,akn},
\ba
\label{pointaction}
S_{{\rm eff}}&=&\frac{T}{2}\,\sum_\omega\ \theta^T(-\omega)\,\big[P(\omega) \big]^{-1}\,
\theta(\omega),\quad \theta=\big( \theta^+_c, \theta^+_s, \theta^-_c, \theta^-_s
\big),\nonumber \\
P(\omega)&=&\frac{4\pi}{|\omega|}\,
\pmatrix{p_{cc}^+ &  p_{cs}^+ & 0 & 0 \cr   p_{cs}^+ &  p_{ss}^+  &  0 & 0 \cr
0  & 0  &  p_{cc}^- &  p_{cs}^-  \cr  0  & 0  &  p_{cs}^- &  p_{ss}^- \cr}, \nonumber \\
p_{cc}^+ &=& \frac{K^{-4}}{v_1+v_2}\,\Big[\frac{ v_{\sigma +}( v_{\rho -} v_{\sigma -} -g_-^2)}{v_1 v_2}+
v_{\rho -} K^2 \Big],\quad
p_{ss}^+=\frac{K^{-4}}{v_1+v_2}\,
\Big[\frac{ v_{\rho +}( v_{\rho -} v_{\sigma -} -g_-^2)}{v_1 v_2}+
v_{\sigma -} K^2 \Big], \nonumber \\
p_{cs}^+ &=&\frac{K^{-4}}{v_1+v_2}\,\Big[\frac{ -g_+( v_{\rho -} v_{\sigma -} -g_-^2)}{v_1 v_2}+
g_- K^2 \Big],\quad p_{cc}^- = \frac{K^{-4}}{v_1+v_2}\,\Big[\frac{ v_{\sigma -}( v_{\rho +} v_{\sigma +} -
g_+^2)}{v_1 v_2}+v_{\rho +} K^2 \Big],  \nonumber \\
p_{ss}^-&=&\frac{K^{-4}}{v_1+v_2}\,\Big[\frac{ v_{\rho -}( v_{\rho +} v_{\sigma +} -g_+^2)}{v_1 v_2}+
v_{\sigma +} K^2 \Big],\quad 
p_{cs}^- =\frac{K^{-4}}{v_1+v_2}\,\Big[\frac{ -g_-( v_{\rho +} v_{\sigma +}-g_+^2)}{v_1 v_2}+
g_+ K^2 \Big], \nonumber \\
v_1^2+v_2^2&=&\Big(v_{\rho +} v_{\rho -} + v_{\sigma +} v_{\sigma -} + 2 g_+ g_-
\Big)/K^2,\quad v_1 v_2 = \Big[(v_{\rho +} v_{\sigma +}- g_+^2 )\,(v_{\rho -} v_{\sigma
-}-g_-^2) \Big]^{1/2}/K^2.
\ea
At high frequency ( or temperature) ,  the inter-edge LRCI can be ignored because
 $ K_0(x) \sim e^{-x}/\sqrt{x}, \quad  {\rm for}\;\; x >> 1$. In this region, it is easy to verify that
$p^+_{cc}=p^+_{ss}=p^-_{cc}=p^-_{ss}=1/K,\quad p^+_{cs}=p^-_{cs}=0 $, and that 
the effective action (\ref{pointaction}) 
depends {\em only} on  $K$. Then, the effective action reduces to that of CLL \cite{imura3}.
At low energy, the inter-edge LRCI is operative, 
and the elements of $P$-matrix depend on energy scale.
In practice,  the terms of order $g_\pm^2$ can be 
neglected. In this approximation the elements of P-matrix simplify.
\ba
p^+_{cc}&\sim&\frac{1}{K \eta_\rho},\quad 
p^+_{ss}\sim\frac{1}{K \eta_\sigma},\quad
p^-_{cc}\sim\frac{\eta_\rho}{K},\quad p^-_{ss}\sim \eta_{\sigma } K,\quad
\eta_\rho\equiv\sqrt{\frac{v_{\rho +}}{v_{\rho -}}},\quad
\eta_\sigma \equiv\sqrt{\frac{v_{\sigma +}}{v_{\sigma -}}}, \nonumber \\
p^+_{cs}&\sim&\frac{g_-}{(\sqrt{v_{\rho +}v_{\rho -} }+
\sqrt{v_{\sigma +}v_{\sigma -} }) K} \ll 1,\quad
p^-_{cs}\sim\frac{g_+}{(\sqrt{v_{\rho +}v_{\rho -} }+\sqrt{v_{\sigma +}v_{\sigma -}
})\,K} \ll 1.
\ea
The explicit expressions of $\eta_\rho$ and $\eta_\sigma$ are ($ w_\parallel\equiv
\sqrt{w_{\uparrow \uparrow} w_{\downarrow \downarrow}},\;\;v_0=\frac{e^2}{\pi \hbar
\epsilon},\quad \Lambda_w= \frac{v_R}{\sqrt{w_\parallel w_\perp}} $ )
\ba
\eta_\rho(\omega)&=&\cases{\Big[1+\xi_\rho \, \ln \frac{\Lambda_w}{|\omega|} 
\Big]^{1/2}, \;\;{\rm for}\;\;
|\omega| < \Lambda_w \cr
1 ,\;\;|\omega| > \Lambda_w \cr},\quad 
\xi_\rho \equiv \frac{ 2 v_0 }{v+v_{{\rm int}}+ v_0\,\ln \sqrt{\frac{w_\parallel
w_\perp}{a_\parallel a_\perp}}}, \nonumber \\
\eta_\sigma(\omega)&=&\cases{ \Big[1 + \xi_\sigma \, \ln \frac{w_\perp}{w_\parallel} \Big]^{1/2}
\equiv \alpha_\sigma,\;\;{\rm for}\;\;
|\omega| < \Lambda_w \cr
1,\;\;{\rm for} \;\; |\omega| > \Lambda_w \cr }
\quad \xi_\sigma \equiv \frac{ v_0 }{ v-v_{{\rm int}}+ v_0 \ln 
\sqrt{\frac{ a_\perp w_\parallel}{a_\parallel w_\perp}}},
\ea
where $v_R$ is the velocity determined by $v_0, v$ and the short range  interactions.
The factor $\eta_\rho(\omega)$  originates from the inter-edge LRCI,  and it  
makes tunneling conductances to decrease faster than any other power law at sufficiently low temperatures
\cite{moon,imura2}.
The factor $\eta_\sigma(\omega)$ characterizes the anomalous spin excitation induced by the
spin dependence of LCRI and is specific to the partially spin
polarized system \cite{leeyang2}. Note that if $w_\perp=w_\parallel$, then 
$\alpha_\sigma = 1$ identically, which
means that the anomalous exponent is entirely due to the spin dependence of inter-edge LRCI. For 
{\em quantum wire} in a  strong magnetic field $\alpha_\sigma-1$ can be estimated to be about 0.1. 
In QH singlet edge states, $\alpha_\sigma-1$ can be significantly large if the short range interaction between
spin-up and spin-down channel is strong enough, as can be seen from the defintion of $\xi_\sigma$
\cite{stability}.

There are two pictures which describe the conductance through the point contact: 
the quasi-particle tunneling (QPT) picture, 
and  the electron tunneling (ET) picture. In the QPT picture, the quasi-particles with opposite chirality
 tunnel through the bulk fractional Hall (FQH) liquid, which is equivalent to the  backscattering by scattering 
potential localized at $x=0$. In ET picture, the electrons tunnel through the depleted region created by the 
negative voltage at the point contact between left and right FQH liquids. 
For spin polarized CLL edge states at
  $\nu=\frac{1}{2 k +1}$, both pictures give the same conductance $ G(T) \sim T^{2/\nu -2}$ at low temperature.
But two pictures give different results for two channel edges, and it is argued that there exists
a crossover between two results  \cite{imura1}. 
To complete the tunneling action we need to specify the scattering potential at $x=0$. The detailed form of the
scattering potential depends on the choice of the tunneling picture. In the next section the quasi-particle
tunneling picture is considered first. 
\section{Quasiparticle Tunneling Picture}
The quasiparticle  scattering potential at the point contact is \cite{imura3}
\be
\label{qpt}
U=-u_{\alpha \beta}\,\big( t^{\alpha \beta}+{\rm H.C} \big),\;\;
 t^{\alpha \beta}=\pmatrix{ e^{i/2(\theta^{+}_c + \theta^{+}_s)} &
 e^{i/2(\theta^{+}_c + \theta^{-}_s)} \cr
e^{i/2(\theta^{+}_c - \theta^{-}_s)} & 
e^{i/2(\theta^{+}_c - \theta^{+}_s)} \cr}, \quad \alpha, \beta=\uparrow,\downarrow.
\ee
The off-diagonal matrix element represents the spin-flip scattering. 

We first consider the high temperature limit where the backscattering is weak, so that 
the perturbative renormalization group treatment is valid.
From (\ref{pointaction}) and (\ref{qpt})
the scaling equations of the scattering amplitudes can be obtained \cite{imura2,zwerger},
\ba
\frac{d \, u_{\uparrow \uparrow}(\Lambda)}{u_{\uparrow \uparrow}(\Lambda)}&=&
\Big[ \frac{p^+_{cc}+p_{ss}^+ +2 p_{cs}^+}{2}-1 \Big]\,\frac{ d \Lambda}{\Lambda},\quad \quad 
\frac{d \, u_{\downarrow \downarrow}(\Lambda)}{u_{\downarrow \downarrow}(\Lambda)}=
\Big[ \frac{p^+_{cc}+p_{ss}^+ -2 p_{cs}^+}{2}-1 \Big]\,\frac{ d \Lambda}{\Lambda} \nonumber \\
\frac{d \, u_{\uparrow \downarrow}(\Lambda)}{u_{\uparrow \downarrow}(\Lambda)}&=&
\frac{d \, u_{\downarrow \uparrow}(\Lambda)}{u_{\uparrow \downarrow}(\Lambda)}=
\Big[ \frac{p^+_{cc}+p_{ss}^-}{2}-1 \Big]\,\frac{ d \Lambda}{\Lambda},
\ea
where $\Lambda$ is the energy cut-off.
The perturbative treatment fails when the renormalized coupling becomes comparable to the cut-off. 
The crossover
temperature which separates the weak and strong coupling regime is 
 $\Big[ \Lambda_1=\big( u_0^3/\Lambda_0 \big)^{1/2}\Big] $\cite{imura2}. 
$u_0, \Lambda_0$ are the bare coupling 
 constants and cut-off, respectively.
 If $\Lambda_1$ is greater than $\Lambda_w$, then
the LRCI does not play any role,  and the   scaling equations are identical
with those of CLL model\cite{imura3}
$G(T) \sim \frac{2}{3} \frac{e^2}{h}- T^{-4/3}.$
If $\Lambda_1$ is smaller than $\Lambda_w$  there are two crossover regions.
For $ \Lambda_1 < \Lambda_w < T $, again the inter-edge Coulomb interaction is irrelevant and 
the temperature dependence of the conductance is identical with that of CLL  model.
For $ \Lambda_1 < T < \Lambda_w $, the renormalized backscattering amplitudes are
\ba
u_{\uparrow \uparrow}(T)&=&u_{\uparrow \uparrow}(\Lambda_w)\,\exp\Bigg[
-\frac{1}{K \xi_\rho}\, \Big[ \sqrt{1+ \xi_\rho \ln \frac{\Lambda_w}{T}}-1 \Big]-
\frac{1}{2 K \alpha_\sigma}\,\ln \frac{\Lambda_w}{T} - \frac{ 4 g_-}{v_R}\,
\big(\ln \frac{\Lambda_D}{T}\big)^{1/2}
\Bigg] \nonumber \\
u_{\downarrow \downarrow}(T)&=&u_{\downarrow \downarrow}(\Lambda_w)\,\exp\Bigg[
-\frac{1}{K \xi_\rho}\, \Big[ \sqrt{1+ \xi_\rho \ln \frac{\Lambda_w}{T}}-1 \Big]-
\frac{1}{2 K \alpha_\sigma}\,\ln \frac{\Lambda_w}{T} + \frac{ 4 g_-}{v_R}\,
\big(\ln \frac{\Lambda_D}{T}\big)^{1/2}
\Bigg] \nonumber \\
u_{\uparrow \downarrow}(T)&=&u_{\uparrow \downarrow}(\Lambda_w)\,\exp\Bigg[
-\frac{1}{K \xi_\rho}\, \Big[ \sqrt{1+ \xi_\rho \ln \frac{\Lambda_w}{T}}-1 \Big]-
\frac{\alpha_\sigma}{2 K}\,\ln \frac{\Lambda_w}{T} \Bigg],
\ea
where $ D \sim v_R /(a_\parallel a_\perp w_\parallel w_\perp)^{1/4} $.
The first term in the exponents comes from the charge sector and is similar to that of
spin polarized Laughlin edge states, while the second term appear only in the spin singlet
egde states. Due to the anomalous factor $\alpha_\sigma$, the non spin-flip and spin-flip 
scattering amplitudes acquire the different temperature dependence.
The third term  in the exponent of $u_{\uparrow \uparrow}(T)$ and $u_{\downarrow \downarrow}(T)$
is due to the direct Zeeman splitting, and it is negligible owing to the factor $g_-/v_R \ll 1$.
Expanding the square root , the conductance 
$(G(T) \sim \nu \frac{e^2}{h}-(\frac{u_{\alpha \beta}}{T})^2 )$ becomes ($K=3$ is substituted).
\ba
G(T) &\sim& \frac{2}{3} \, \frac{e^2}{h}- 
\Big(\frac{T}{\Lambda_w} \Big)^{-\frac{4}{3}+(1/\alpha_\sigma -1)/3}\,e^{\frac{\xi_\rho}{24}\,
\ln^2 \frac{\Lambda_w}{T} } ,\quad \mbox{no spin-flip} \nonumber \\
G(T) &\sim& \frac{2}{3} \, \frac{e^2}{h}- 
\Big(\frac{T}{\Lambda_w} \Big)^{-\frac{4}{3}+(\alpha_\sigma -1)/3}\,e^{\frac{\xi_\rho}{24}\,
\ln^2 \frac{\Lambda_w}{T} } ,\quad \mbox{ spin-flip}.
\ea
The anomalous exponent $\alpha_\sigma$
{\em enhances} the non spin-flip backscattering amplitude and {\em suppresses}
the spin-flip backscattering.

At low temperaure $ T < \Lambda_1 $ where the backscattering becomes very strong
 the dominant transport process is the tunneling 
between minima of the quasi-particle scattering potential. 
Using the duality mapping in the dilute instanton gas approximation(DIGA) \cite{schmid}, the
original model can be mapped into the model with weak potential \cite{imura2,imura3}.
The $\theta^-_c$  which does not appear in the QPT term has to be integrated out before performing
the DIGA. The resulting dual effective action is
\be
S_{{\rm DIGA}}=\frac{T}{2}\,\sum_\omega\,\frac{|\omega|}{\pi }\,\tilde{\theta}^T\,
\pmatrix{ p^+_{cc} &  p^+_{cs} & 0 \cr
        p^+_{cs} &  p^+_{ss} & 0 \cr
	0        &      0    & p^-_{ss} \cr}\, \tilde{\theta}-
	2\,y_j\,\cos\Big( \sum_I\,C_{Ij}\,\tilde{\theta}_I \Big),\quad
\vec{\tilde{\theta}}=\big(\tilde{\theta}_c^+, \tilde{\theta}_s^+, \tilde{\theta}_s^- \big),
\ee
where $y_j$ is the instanton fugacity of the $j$th species. $C_{Ij}$ is the instanton transition
matrix element in the lattice of potential minima \cite{imura1}. If there is no spin-flip
scattering, the {\em least irrelevant}
allowed transition vectors are
\be
\vec{C}_{nsf,1}=(2,0,0), \quad 
\vec{C}_{nsf,2}=(0,2,0), \quad 
\vec{C}_{nsf,3}=(1,1,0), \quad
\vec{C}_{nsf,4}=(1,-1,0).
\ee
When the spin-flip scattering is present,  the allowed vectors are \cite{imura3}
\ba
\vec{C}_{sf,1}&=&(2,0,0), \quad \vec{C}_{sf,2}=(0,2,0), \quad \vec{C}_{sf,3}=(0,0,2), \quad
\vec{C}_{sf,4}=(1,1,1),\nonumber \\
\vec{C}_{sf,5}&=&(1,1,-1),\quad \vec{C}_{sf,6}=(1,-1,1), \quad \vec{C}_{sf,7}=(1,-1,-1).
\ea
The scaling equations of fugacities are
\be
\frac{ d ( y_j /\Lambda ) }{y_j/\Lambda}=\bigg[ \frac{1}{2} \Big(
\frac{ (C^{c+}_j)^2\,p^{+}_{ss} + (C^{s+}_j)^2\,p^{+}_{cc} + 2 C_j^{c+} C_j^{s+}\,p^+_{cs}}{
p^+_{cc} p^+_{ss}-(p^+_{cs})^2} +\frac{ (C_j^{s-})^2}{p^-_{ss}} \Big) -1 \bigg]\,\frac{d
\Lambda}{\Lambda}.
\ee
In the region $ \Lambda_w < T < \Lambda_1 $ LRCI has no effect on the RG equation, where
R. G equation reduces to
\be
\frac{ d ( y_j /\Lambda ) }{y_j/\Lambda}=\bigg[ \frac{K}{2} \Big(
(C_j^{c+})^2+ (C_j^{s+})^2 + (C_j^{s-})^2 \Big) -1 \bigg]\,\frac{d\Lambda}{\Lambda}.
\ee
Substituting the least irrelevant $\vec{C}_{nsf,3}$ and $\vec{C}_{nsf,4}$ for non spin- flip case, 
and $\vec{C}_{sf,4}$, $\vec{C}_{sf,5}$, $\vec{C}_{sf,6}$, $\vec{C}_{sf,7}$
 for spin flip case in $K=3, \nu=2/3$ state, we find $
G(T) \sim T^4$ and $ G(T) \sim T^7$, respectively. The above results coincide with those of Imura and Nagaosa
\cite{imura3} obtained within CLL theory.
The spin conductance is obtained with the choice of 
$\vec{C}_{nsf,2}=\vec{C}_{sf,2}(0,2,0)$ and $\vec{C}_{sf,3}=(0,0,2)$, 
both of them giving the same result  $G_s(T) \sim T^{10}$.

At lower temperature $ T < \Lambda_w $, the LCRI is operative.  In the temperature range
$ \Lambda_w \,e^{-\frac{1}{\xi_\rho}} < T < \Lambda_w$ the charge conductance is given by,
\be
G(T) \sim \Big(\frac{y_j(T)}{T} \Big)^2
\sim T^{K\big[(C_j^{c+})^2+\alpha_\sigma \, (C_j^{s+})^2+\alpha_\sigma^{-1}\,(C_j^{s-})^2 \big]-2}
\,e^{-\frac{K}{4}\, \xi_\rho\, (C_j^{c+})^2\,\ln^2\frac{\Lambda_w}{T}}.
\ee
Substituting the same instanton transition vector as the above  CLL case  
we find the anomalous contribution from the spin sector
{\em suppresses} the charge tunneling conductance by 
$T^{K(\alpha_\sigma -1)}$ for non spin-flip case, and 
$T^{K(\alpha_\sigma + \alpha_\sigma^{-1}-2)}$ for  spin-flip case  compared to the conductance 
 obtained  within CLL theory.
In addition, the anomalous contributions are  {\em non-universal} because they depend on the 
detailed shape of confining edge potential.

 In the spin conductance channel
$\vec{C}_{sf,2}=(0,2,0)$ and $\vec{C}_{sf,3}=(0,0,2)$ are degenerate at $ T > \Lambda_w $,
while at $ \Lambda_w \,e^{-\frac{1}{\xi_\rho}} < T < \Lambda_w$, $\vec{C}_{sf,3}$ mode is more dominant.
The spin conductance in  the non spin-flip channel is {\em suppressed }
by the anomalous exponent $\alpha_\sigma$
, but the spin conductance in  the spin-flip channel
is {\em enhanced} compared to the CLL case.
\be
G_s(T) \sim T^{10}\cdot T^{-12(1-\alpha_\sigma^{-1})}, \quad  \mbox{Spin-Flip Channel}.
\ee
The enhancement factor is larger for larger $K$, namely at lower filling fraction.
Note that the direct Coulomb 
suppression  factor $\eta_\rho$  from the charge sector is absent
in the spin channel  $\vec{C}_{sf,2}$ and $\vec{C}_{sf,3}$ , which is an indictation of spin charge separation.
At very low temperature 
$ T < e^{-\frac{1}{\xi_\rho}}\,\Lambda_w $ the charge conductance is
\be
G(T) \sim \exp\Big[-\frac{K \sqrt{\xi_\rho}}{3}\, 
\big(\ln \frac{\Lambda_w}{T}\big)^{3/2}\Big]\,
T^{K(\eta_\sigma + \eta_\sigma^{-1})-2},
\ee
which decreases faster than any other power law \cite{moon,imura2}. The anomalous exponent makes the charge
conductance to decrease even faster, although its contribution is subleading.
\section{Electron Tunneling Picture}
At low temperature where the quasi-particle tunneling is irrelevant it is more appropriate to
start from the electron tunneling (ET) picture. 
The operator describing the electron tunneling from the left to the right edge is
\be
\Gamma=\sum_{I,J=\uparrow,\downarrow}\, \gamma_{I J}\,
\Psi^{(L)}_I(x=0) \,\Psi^{(R) \dag}_{J}(x=0) + {\rm H.C}.
\ee
$\Psi$ is the generalized electron operator \cite{imura3,elop}$
\Psi^{(L)}_I(x=0)=\sum_{n=0,1}\, c_n \, \exp\Big[ i n \theta^{(L)}_{-I} + i ( K-n) 
\theta^{(L)}_I \Big]$.
The conductance in the electron tunneling picture is determined by the scaling dimension of the
$\gamma_{I J}$ \cite{imura1}. 
If the width of QH bar is much greater than 
the separation between the left and right condensate, and if the temperature is higher than 
$\Lambda_w$ the left and right edges can be assumed to be parallel and infinitely long \cite{imura2}. 
The effective action for the electron tunneling in this regime is formally identical with 
that of QPT Eq.(\ref{pointaction})  if $w_{s s^{\prime}} $ are replaced with $ d_{s s^{\prime}}$ 
, even though their physical origins are different. 
The effect of  anomalous coupling in spin sector
$\alpha_\sigma - 1$  is more pronounced in  ET picture than in QPT picture 
because the inter-edge LRCI length scale
is much shorter in ET picture ($ d \ll w, \quad \Lambda_d \gg \Lambda_w $).
The scaling equation is 
\be
\label{etscale}
\frac{ d \gamma_{IJ} / \Lambda}{\gamma_{IJ} /\Lambda}=
\Bigg[\frac{2}{K}\,\Big[ \ell^2\,p^+_{cc,{\rm ET}}+ m^2\,p^+_{ss,{\rm ET}}+n^2\,p^-_{ss,{\rm ET}}+
2 \ell m \, p^+_{cs,{\rm ET}} \Big]-1\Bigg]\,\frac{d \Lambda}{\Lambda},
\ee
where $p^{+,-}_{c,s,{\rm ET}}$'s are the P-matrix elements in the ET picture. The set of rational numbers
$\vec{C}_{{\rm ET}}=(\ell, m, n)$ characterize the various terms in the electron tunneling operator 
$(\Gamma=\sum e^{i \ell \theta^+_{cc,{\rm ET}}+i m \theta^+_{ss,{\rm ET}} 
+ i n \theta^-_{ss,{\rm ET}} } )$.
The least irrelevant set of $\vec{C}_{{\rm ET}}$ are $(3/2, 1/2, 0)$ and $(3/2, 0, 1/2)$, and they 
correspond to the non spin-flip and spin-flip tunneling, respectively.
At $ \Lambda_1 > T > \Lambda_d=v_R/d $ LRCI is effectless, and the conductance becomes
$ G(T) \sim T^{4/3}$, which agrees with the result of CLL theory\cite{imura1}.
In the temperature range $ \Lambda_d e^{-\frac{1}{\xi_{\rho,ET}}} < T < \Lambda_d $ 
the LRCI is effective, and the condutance behaves like
\ba
G_{{\rm ET}}(T) &\sim& \gamma_{\uparrow \uparrow}^2 (T)  \sim  \gamma_{\downarrow \downarrow}^2 (T) 
\sim T^{4/3 + 4/3(\alpha_\sigma^{-1}-1)}\,
e^{\frac{3}{4}\,\xi_{\rho,{\rm ET}}\,\ln^2 \frac{\Lambda_d}{T}}, \quad \mbox{No spin-flip} \nonumber \\
G_{{\rm ET}}(T) &\sim& \gamma_{\uparrow \downarrow}^2(T) \sim T^{4/3 + 4/3(\alpha_\sigma-1)}\,
e^{\frac{3}{4}\,\xi_{\rho,{\rm ET}}\,\ln^2 \frac{\Lambda_d}{T}}, \quad \mbox{spin-flip}.
\ea
The anomalous interaction in spin sector {\em enhances} the  non spin-flip
tunneling conductance, while it {\em suppresses} the spin-flip tunneling conductance.
This should be compared with the result obtained in QPT picture with DIGA , where the 
anomalous interaction in spin sector {\em suppresses} {\em both} 
the non spin-flip tunneling 
conductance {\em and} the spin-flip tunneling conductance.

Below the temperature where the right hand side of scaling equation (\ref{etscale}) vanishes  ET becomes
relevant, and the conductance rises with the decreasing temperature.
But at temperature below $\Lambda_w$ the edges extended to the left and right need to be taken into account
\cite{imura2}, and the separate treatment is required.
Following the treatment in \cite{imura2} we start with the QPT model with LRCI and
we discard the $x > 0$ segment. This is equivalent to imposing  the 
 constraint  $\delta(\theta^+_I(x=0))$, then
it is convenient to formulate the problem in terms of the dual $\theta^-_I$ field \cite{kanefisher}.
Including the contribution from the right branch also we get the electron tunneling action valid at temperature 
$T < \Lambda_w$.
\be
S_{{\rm ET}}=\frac{T}{2}\,\sum_{i=L,R}\,\sum_\omega \, 
\big( \theta^-_{c i}, \theta^-_{s i} \big)\, \big[P_2 \big]^{(-1)} \,
 \pmatrix{  \theta^-_{c i} \cr \theta^-_{s i} \cr},\quad
P_2 \sim \frac{ 8 \pi}{|\omega|}\pmatrix{ \eta_\rho &  \frac{g_+}{v_\rho + v_\sigma} \cr
\frac{g_+}{v_\rho + v_\sigma} &  \eta_\sigma }.
\ee
The electron tunneling term is $ \sum_{I,J}\,\big( \Psi_{L I}\Psi^{\dag}_{RJ} 
+ {\rm H.C} \big)$.
Neglecting the term of order $g_+$, the scaling equation is
\be
\frac{ d t_{\uparrow \uparrow} / \Lambda}{t_{\uparrow \uparrow} / \Lambda}=
\frac{ d t_{\uparrow \downarrow} / \Lambda}{t_{\uparrow \downarrow} / \Lambda}=
\Big[ \frac{1}{2}\,\big( 9\, \eta_\rho + \eta_\sigma \big) -1 \Big]\,\frac{d
\Lambda}{\Lambda}.
\ee
Integrating the above scaling equation from $T$ to $\Lambda_w$, we find that
$G(T) \sim T^{7+\alpha_\sigma}$ for $ e^{-\frac{1}{\xi_\rho}}\,\Lambda_w < T < \Lambda_w $, and 
at very low temperature $ T < e^{-\frac{1}{\xi_\rho}}\,\Lambda_w $,
$G(T) \sim \frac{1}{T^2}\,e^{-3 \sqrt{\xi_\rho}\,\ln^{3/2} \frac{\Lambda_w}{T}}\,T^{\alpha_\sigma}$.
Compared with the results of the QPT, the conductance obtained in ET picture at $T < \Lambda_w$  
decreases {\em faster} with temperature. 
\section{Summary}
In summary, we studied the effect of the anomalous interaction in the spin sector of FQH spin 
singlet edge states on the charge and spin tunneling conductances. 
The anomalous interaction  is induced by the interplay of Zeeman 
splitting and the LRCI between edges of opposite chirality.
The anomalous interaction in spin sector can revealed  as the anomalous exponent $\alpha_\sigma$ which is always 
greater than unity. The conductances are enhanced or suppressed depending on the kinds of scattering at 
the point contact and the mechanism of tunneling (QPT, ET).
The above effect may be observed experimentally in the double layer  Hall system \cite{eisenstein}
at low magnetic field in the temperature range $ \Big[ T \le 10 {\rm mK} \Big ]$ \cite{moon}. 
It is also relevant to the recent NMR measurements of QH edges \cite{nmr}.

The author is grateful to Hangmo Yi for useful comments.



\begin{references}
\bibitem{wen} X.  G. Wen, Phys. Rev. B {\bf 41}, 12838 (1990);
X. G. Wen, Int. Journ. Mod. Phys. B {\bf 6},1711 (1992).
\bibitem{experiments} F. P. Milliken, C. P. Umbach, and R. A. Webb, Sol. St. Comm. {\bf 97}, 309
(1996); A. M. Chang, L. N. Pfeiffer, and K. W. West, Phys. Rev. Lett. {\bf 77}, 2538 (1996);
K. Moon, {\it et al.}, Phys. Rev. Lett. {\bf 71}, 4381 (1992).
\bibitem{moon} K. Moon and S. M. Girvin, Phys. Rev. B {\bf 54}, 4448 (1996).
\bibitem{imura2} K. Imura and N. Nagaosa, Sol. Stat. Com. {\bf 103}, 663 (1997).
\bibitem{brey} M. Franco and L. Brey, Phys. Rev. Lett. {\bf 77}, 1358 (1996).
\bibitem{fendley} P. Fendley, A. W. W. Ludwig, and H. Saleur, Phys. Rev. B {\bf 52}, 8934 (1995).
\bibitem{spin-charge} H. C. Lee and S. -R. Eric Yang, Phys. Rev. B {\bf 56}, R15529 (1997).
\bibitem{schulz} H. J. Schulz, Phys. Rev. Lett. {\bf 71}, 1864 (1993).
\bibitem{leeyang2} H. C. Lee and S. -R. Eric Yang, Phys. Rev. B {\bf 57}, R4249 (1998).
\bibitem{imura3} K. Imura and N. Nagaosa, Phys. Rev. B  {\bf 57}, R6826 (1998).
\bibitem{eisenstein} J. P. Eisenstein, H. L. Stormer, L. N. Pfeiffer, and K. W. West, Phys. Rev. Lett. 
{\bf 62}, 1540 (1989); Phys. Rev. B {\bf 41}, 7910 (1990).
\bibitem{macdonald} I. A. MacDonald and F. D. M. Haldane, Phys. Rev. B {\bf 53}, 15845 (1996).
\bibitem{wenzee} X. G. Wen and A. Zee, Phys. Rev. B {\bf 46}, 2290 (1992).
\bibitem{kfp} C. L. Kane, M. P. A. Fisher, and J. Polchinski, 
Phys. Rev. Lett. {\bf 72}, 4129 (1994).
\bibitem{zeeman} The amount of the edge spin-splitting must be calculated including many body interactions;
J. Dempsey, B. Y. Gelfand, and B. I. Halperin, Phys. Rev. Lett. {\bf 70}, 3639 (1993);
S. -R. Eric Yang, A. H. MacDonald, and M. D. Johnson, Phys. Rev. Lett. {\bf 71}, 3194 (1993).
\bibitem{note1} We consider  general cases $\hat{K}
=(K/K/0)$. For $\nu=2/3$, $K=3$. 
\bibitem{kanefisher} C. Kane and M. P. A. Fisher, Phys. Rev. B {\bf 46}, 15233 (1992).
\bibitem{akn} A. Furusaki and N. Nagaosa, Phys. Rev. B {\bf 47}, 3827 (1993).
\bibitem{stability}We assume that the stability condition \cite{imura3} is satisfied.
\bibitem{imura1} K. Imura and N. Nagaosa, Phys. Rev. B {\bf 55}, 7690 (1997).
\bibitem{zwerger} M. P. A. Fisher and W. Zwerger, Phys. Rev. B {\bf 32}, 6190 (1985).
\bibitem{schmid} A. Schmid, Phys. Rev. Lett. {\bf 51}, 1506 (1983).
\bibitem{elop}  X. G. Wen, Phys. Rev. B {\bf 44}, 5708 (1991).
\bibitem{nmr} K. R. Wald, {\it et al.}, Phys. Rev. Lett. {\bf 73}, 1011 (1994);
Dixon, {\it et al.}, cond-mat/9701039.
\end{references}
\end{document}